\documentclass{elsart}%
\usepackage{amsfonts}
\usepackage{amsmath}
\usepackage{amssymb}
\usepackage{graphicx}
\usepackage{graphicx}
\usepackage{psfig}
\usepackage{amsmath}%
\begin{document}
%
%TCIMACRO{\TeXButton{Begin frontmatter}{\begin{frontmatter}}}%
%BeginExpansion
\begin{frontmatter}%
%EndExpansion
%

%TCIMACRO{\TeXButton{Title}{\title
%{On the ground state energy of a gas of interacting polarons in a magnetic field}%
%}}%
%BeginExpansion
\title
{On the ground state energy of a gas of interacting polarons in a magnetic field}%
%EndExpansion
%

%TCIMACRO{\TeXButton{Author}{\author{K. Putteneers}\ead
%{katrijn.putteneers@ua.ac.be},
%\author{F. Brosens}\ead{fons.brosens@ua.ac.be},
%\author{S.N. Klimin}\ead{sergei.klimin@ua.ac.be},
%\author{J.T. Devreese}\ead{jozef.devreese@ua.ac.be}}}%
%BeginExpansion
\author{K. Putteneers}\ead{katrijn.putteneers@ua.ac.be},
\author{F. Brosens}\ead{fons.brosens@ua.ac.be},
\author{S.N. Klimin}\ead{sergei.klimin@ua.ac.be},
\author{J.T. Devreese}\ead{jozef.devreese@ua.ac.be}%
%EndExpansion
%

%TCIMACRO{\TeXButton{Address}{\address
%{Departement Fysica, Universiteit Antwerpen, Universiteitsplein 1,  B-2610 Antwerpen}%
%}}%
%BeginExpansion
\address
{Departement Fysica, Universiteit Antwerpen, Universiteitsplein 1,  B-2610 Antwerpen}%
%EndExpansion
%

%TCIMACRO{\TeXButton{Begin abstract}{\begin{abstract}} }%
%BeginExpansion
\begin{abstract}
%EndExpansion
The ground-state energy of a three-dimensional polaron gas in a magnetic field
is investigated. An upper bound for the ground-state energy is derived within
a variational approach which is based on a many-body generalization of the
Lee-Low-Pines transformation. The basic contributing ingredients found are the
ground-state energy and the static structure factor of the homogeneous
electron gas in a magnetic field. Both these quantities are derived in the
Hartree-Fock approximation. The resulting ground-state energy of the polaron
gas is analyzed as a function of the electron density and of the magnetic
field strength.
%TCIMACRO{\TeXButton{End abstract}{\end{abstract}}}%
%BeginExpansion
\end{abstract}%
%EndExpansion
%

%TCIMACRO{\TeXButton{Begin keywords}{\begin{keyword}} }%
%BeginExpansion
%EndExpansion
%

%TCIMACRO{\TeXButton{End frontmatter}{\end{frontmatter}}}%
%BeginExpansion
\end{frontmatter}%
%EndExpansion

\section{Introduction}

Increasing interest in the polaron gas problem is stimulated by intense
experimental investigations of polar materials (see, for review,
Refs.~\cite{alex,Polarons}). The electron-phonon interaction is manifested, in
particular, in experiments on the cyclotron resonance in bulk \cite{McCombe68}
and quasi-2D \cite{Chang88,Cheng91,Peeters92} systems, giving evidence for a
resonant magneto-polaron effect \cite{Larsen1964}. Cyclotron-resonance
measurements on semiconductor quantum wells with high electron density
\cite{Poulter2001,F2004a,Faugeras2004} reveal that the CR lines are shifted
and split as compared to those in low-density quantum wells. In Ref.
\cite{PRB2003}, these effects are interpreted within the framework of the
polaron concept, which thus remains valid for high-electron density systems.
In this connection, the investigation of the ground-state properties of an
interacting polaron gas is an important ingredient.

{\normalsize In the present work, we analyze an interacting polaron gas at
}$T=0${\normalsize . However, the obtained results can also be applied to a
polaron gas at nonzero temperature if the thermal energy is small as compared
to both the LO-phonon energy and the Fermi energy (}$k_{B}T\ll\hbar
\omega_{\mathrm{LO}}${\normalsize , }$k_{B}T\ll E_{F}${\normalsize ). }An
upper bound for the ground-state energy of a polaron gas in an applied
homogeneous magnetic field is derived using a many-body generalization
\cite{LDB77} of the familiar Lee-Low-Pines transformation \cite{LLP} for the
polaron. The Hamiltonian of the polaron gas is given by {%
\begin{align}
H  &  =\frac{1}{2m}\sum_{\sigma}\int d^{3}r\Psi_{\sigma}^{\dag}(\mathbf{r}%
)\left(  \frac{\hbar}{i}\mathbf{\nabla}+\frac{e}{c}\mathbf{A}(\mathbf{r}%
)\right)  ^{2}\Psi_{\sigma}(\mathbf{r})\nonumber\\
&  +\frac{1}{2}\sum_{\sigma\sigma^{\prime}}\int d^{3}r\int d^{3}r^{\prime}%
\Psi_{\sigma}^{\dag}(\mathbf{r})\Psi_{\sigma^{\prime}}^{\dag}%
(\mathbf{r^{\prime}})v(\mathbf{r}-\mathbf{r^{\prime}})\Psi_{\sigma^{\prime}%
}(\mathbf{r^{\prime}})\Psi_{\sigma}(\mathbf{r})\nonumber\\
&  +\sum_{\mathbf{k}}\hbar\omega_{\mathrm{LO}}a_{\mathbf{k}}^{\dag
}a_{\mathbf{k}}+\sum_{\mathbf{k}\sigma}\left(  V_{k}^{\ast}a_{\mathbf{-k}%
}^{\dag}+V_{k}a_{\mathbf{k}}\right)  \int d^{3}r\Psi_{\sigma}^{\dag
}(\mathbf{r})e^{i\mathbf{k}\cdot\mathbf{r}}\Psi_{\sigma}(\mathbf{r}%
)\nonumber\\
&  +\sum_{\sigma}\int d^{3}r\Psi_{\sigma}^{\dag}(\mathbf{r})g\mu_{B}%
B\sigma\Psi_{\sigma}(\mathbf{r}).
\end{align}
where }$a_{\mathbf{k}}^{\dag}$ and $a_{\mathbf{k}}$ are the creation and
annihilation operators for the longitudinal optical (LO) phonons with
frequency $\omega_{\mathrm{LO}}$, $g$ is the effective Land\'{e} factor,
$\mu_{B}$ is the Bohr magneton, $m$ is the band mass, and $\Psi_{\sigma}%
^{\dag}(\mathbf{r})$ and $\Psi_{\sigma}(\mathbf{r})$ are the fermion creation
and annihilation field operators for the{ charge carriers} in position{
$\mathbf{r}$ with spin projection $\sigma=\pm1/2$ along the }${z}$ axis{ which
is chosen in the direction of the magnetic field }$\mathbf{B}$. The
electron-phonon interaction strength is described by the amplitude
\begin{equation}
V_{k}=-i\hbar\omega_{\mathrm{LO}}\left(  \frac{4\pi\alpha}{Vk^{2}}\right)
^{1/2}\left(  \frac{\hbar}{2m\omega_{\mathrm{LO}}}\right)  ^{1/4}%
,\,\alpha=\frac{1}{2}\frac{e^{2}}{\hbar\omega_{\mathrm{LO}}}\left(  \frac
{1}{\epsilon_{\infty}}-\frac{1}{\epsilon_{0}}\right)  \left(  \frac
{2m\omega_{\mathrm{LO}}}{\hbar}\right)  ^{1/2},
\end{equation}
{where }$\alpha$ is the dimensionless electron-phonon coupling constant with
the high-frequency and static dielectric constants $\varepsilon_{\infty}$ and
$\varepsilon_{0}$, and $v(\mathbf{r})$ is the Coulomb interaction between the
electrons, with the Fourier transform%
\begin{equation}
v(k)=\frac{1}{V}\frac{4\pi e^{2}}{\varepsilon_{\infty}k^{2}}%
\end{equation}
where $V$ is the crystal volume.

In Ref. \cite{LDB77}, the ground-state energy was studied for a polaron gas in
the absence of a magnetic field by introducing the following unitary phonon
translation operator, which is the many-particle generalization of the
Lee-Low-Pines transformation \cite{LLP}:%
\begin{equation}
U=\exp\left[  {\sum_{\sigma}\int d^{3}r\Psi_{\sigma}^{\dag}(\mathbf{r}%
)}\left(  \sum_{\mathbf{k}}\left(  f_{\mathbf{k}}a_{\mathbf{k}}e^{i\mathbf{k}%
\cdot\mathbf{r}}-f_{\mathbf{k}}^{\ast}a_{\mathbf{k}}^{\dag}e^{-i\mathbf{k}%
\cdot\mathbf{r}}\right)  \right)  {\Psi_{\sigma}(\mathbf{r})}\right]  .
\label{trafo}%
\end{equation}
By applying exactly the same unitary transformation for the polaron gas in a
the presence of a magnetic field and calculating the expectation value
$\left\langle \Psi_{GT}\left\vert U^{-1}HU\right\vert \Psi_{GT}\right\rangle $
{in a trial ground state $\left\vert \Psi_{GT}\right\rangle $}$=\left\vert
vac\right\rangle \cdot\left\vert \Psi_{el}\right\rangle ${ which is the
product of the phonon vacuum and an electron state, we found an upper bound to
the ground-state energy}%
\begin{align}
E  &  =NE_{B}-\frac{1}{2}N\sum_{\mathbf{k}}v(k)\left(  1-S(\mathbf{k})\right)
+N\sum_{\mathbf{k}}\left(  \hbar\omega_{\mathrm{LO}}S(\mathbf{k})+\frac
{\hbar^{2}k^{2}}{2m}\right)  f_{\mathbf{k}}f_{\mathbf{k}}^{\ast}\nonumber\\
&  -N\sum_{\mathbf{k}}(V_{k}f_{\mathbf{k}}^{\ast}+V_{k}^{\ast}f_{\mathbf{k}%
})S(\mathbf{k}). \label{erg}%
\end{align}
{where $E_{B}$ is the energy of a gas of non-interacting electrons in an
applied magnetic field }$B${, and} $S(\mathbf{k})$ is the static structure
factor of a homogeneous electron gas in a magnetic field. Minimizing
{(\ref{erg}) with respect to $f_{\mathbf{k}}^{\ast}$ we obtained the same
formal expression for the optimal values of }$f_{\mathbf{k}}$ {as in
\cite{LDB77}: }%
\begin{equation}
f_{\mathbf{k}}=\frac{V_{\mathbf{k}}}{\hbar\omega_{\mathrm{LO}}+\frac{\hbar
^{2}k^{2}}{2mS\left(  \mathbf{k}\right)  }}.
\end{equation}
This results in {the following expression for the variational ground-state
energy per particle:
\begin{equation}
E_{0}=E_{B}-\frac{1}{2}\sum_{\mathbf{k}}v(k)\left[  1-S\left(  \mathbf{k}%
\right)  \right]  -\sum_{\mathbf{k}}\frac{\left\vert V_{k}\right\vert
^{2}S(\mathbf{k})}{\hbar\omega_{\mathrm{LO}}+\frac{\hbar^{2}k^{2}%
}{2mS(\mathbf{k})}}. \label{grondperdeel}%
\end{equation}
}

The ground-state energy (\ref{grondperdeel}) is structurally similar to that
obtained in \cite{LDB77} in the absence of a magnetic field. The basic
ingredients $E_{B}$ and $S(\mathbf{k)}$ are different though, because of the
magnetic field. The static and dynamic structure factors can be calculated
using various approximations (see, e. g., Ref. \cite{SF1} for a confined 3D
electron gas and Refs. \cite{SF2,SF3,SF4} for a 2D electron gas).

In the present paper, we calculate $E_{B}$ and $S(\mathbf{k})$ in the
Hartree-Fock approximation, which is variational in nature. The contribution
$E_{B}$ is treated in Section \ref{EB}. The static structure factor
$S(\mathbf{k})$ will be discussed in Section \ref{SF}. The upper bound to the
ground-state energy per particle, taking both contributions into account, is
discussed in Section \ref{GSE}. For the numerical calculations and
illustrations the degenerate polar semiconductor GaAs is used (Table 1).

\newpage

\begin{center}
\textbf{Table 1. Parameters and material constants of GaAs which are used for
the calculations.}

\medskip%

\begin{tabular}
[c]{|c|c|c|c|c|c|}\hline
$m/m_{e}$ & $\hbar\omega_{\mathrm{LO}}$ (meV) & $\varepsilon_{\infty}$ &
$\varepsilon_{0}$ & $\alpha$ & $g$\\\hline
0.0653 \cite{Pfeffer} & 36.3 \cite{Adachi} & 10.89 \cite{Adachi} & 13.18
\cite{Adachi} & 0.068 \cite{kart} & $-${\normalsize 0.44 \cite{Oestreich}%
}\\\hline
\end{tabular}

\end{center}

\section{Ingredients contributing to the ground-state energy}

\subsection{Energy of non-interacting electrons in a magnetic field}

\label{EB}

The energy of non-interacting electrons in a magnetic field is calculated
starting from the energy for a single particle in a magnetic field and
applying the Pauli principle. Expressing that the number of particles $N$
equals the number of occupied states we arrive at the equation
\begin{equation}
\nu\sum_{\sigma=\pm\frac{1}{2}}\sum_{n=0}^{\infty}\sum_{k_{z}}\Theta\left(
E_{F}-\hbar\omega_{c}\left(  n+\frac{1}{2}\right)  -g\mu_{B}B\sigma
-\frac{\hbar^{2}k_{z}^{2}}{2m}\right)  =N,\label{pauli}%
\end{equation}
where $\nu$ is the degeneration factor of each Landau level, $k_{z}$
represents the wave vector component along the magnetic field, $\omega_{c}$ is
the cyclotron frequency and $E_{F}$ stands for the Fermi energy. {The function
}$\Theta\left(  x\right)  $ is defined as
\begin{equation}
\Theta\left(  x\right)  =\left\{
\begin{array}
[c]{c}%
1\text{ if }x\text{ is true}\\
0\text{ if }x\text{ is false}%
\end{array}
\right.  .
\end{equation}
By summing all one-electron energies obeying the Pauli principle one obtains
\begin{equation}
E_{B}=\frac{m\omega_{c}}{2\pi^{2}\hbar n_{0}}\sum_{\sigma=\pm\frac{1}{2}}%
\sum_{n=0}^{n_{\max}(\sigma)}\left[  \left(  \hbar\omega_{c}\left(  n+\frac
{1}{2}\right)  +g\mu_{B}B\sigma\right)  \kappa_{n,\sigma}+\frac{\hbar
^{2}\kappa_{n,\sigma}^{3}}{6m}\right]  ,
\end{equation}
{
\begin{equation}
{\kappa_{n,\sigma}=\sqrt{\frac{2m}{\hbar^{2}}\left(  E_{F}-\hbar\omega
_{c}\left(  n+\frac{1}{2}\right)  -g\mu_{B}B\sigma\right)  },}%
\end{equation}
where $n_{0}\equiv N/V$ }is the electron density and $\kappa_{n,\sigma}$ is
the maximal wave vector in $z$-direction of an electron with the spin
projection $\sigma$ in the $n$-th Landau level. The integer{ }%
\begin{equation}
n_{\max}\left(  \sigma\right)  =\left[  \frac{E_{F}-g\mu_{B}B\sigma}%
{\hbar\omega_{c}}-\frac{1}{2}\right]
\end{equation}
{denotes the number of fully occupied Landau levels} with the spin projection{
}$\sigma${. The energy contributions clearly depend on the electron density
and on the magnitude of the applied magnetic field via the Fermi energy and
the number of occupied Landau levels.}

\subsection{Structure factor}

\label{SF}

Within the Hartree-Fock approximation, the expression for the static structure
factor is calculated based on the second quantization form
\begin{align}
S(\mathbf{k})  &  =1+\frac{1}{N}\sum_{\sigma\sigma^{\prime}}\int d^{3}r\int
d^{3}r^{\prime}e^{i\mathbf{k}\cdot(\mathbf{r}-\mathbf{r}^{\prime})}\nonumber\\
&  \times\left\langle \Psi_{el}\left\vert \Psi_{\sigma}^{\dag}(\mathbf{r}%
)\Psi_{\sigma^{\prime}}^{\dag}(\mathbf{r^{\prime}})\Psi_{\sigma^{\prime}%
}(\mathbf{r^{\prime}})\Psi_{\sigma}(\mathbf{r})\right\vert \Psi_{el}%
\right\rangle .
\end{align}
The calculations were performed with periodic boundary conditions on a cube
with volume $V,$ {and the limit }$V\rightarrow\infty$ {is taken keeping the
density }$n_{0}$ {constant. Writing out the field operators in the energy
representation and applying Wick's theorem, the structure factor is expressed
in term}s of integrals over wave vectors and summations over eigenstates of
electrons in a magnetic field. Using the orthogonality relations for the
eige{nstates one is left with an exchange term which involves integrals of the
form \cite{grad} {%
\begin{equation}
\int\limits_{-\infty}^{\infty}dx\text{ }e^{-x^{2}}H_{m}(x+\alpha)H_{n}%
(x+\beta)=2^{n}\sqrt{\pi}m!\beta^{n-m}L_{m}^{n-m}(-2\alpha\beta)\text{ for
}m<n
\end{equation}
where $H_{n}(x)$ is the Hermite polynomial of the $n$-th order and where
$L_{j}^{k}(x)$ is an Laguerre polynomial.}}

{{The resulting expression for the static structure factor of an electron gas
in a magnetic field is{%
\begin{align}
S(\mathbf{k})  &  =1-\frac{1}{(2\pi l_{B})^{2}n_{0}}e^{-\frac{l_{B}^{2}%
k_{\bot}^{2}}{2}}\sum_{\sigma}\sum_{n}^{n_{\max}(\sigma)}\sum_{n^{\prime}%
}^{n_{\max}(\sigma)}\frac{n_{<}!}{n_{>}!}\left(  \frac{l_{B}^{2}k_{\bot}^{2}%
}{2}\right)  ^{n_{>}-n_{<}}\nonumber\\
&  \times\left[  L_{n_{<}}^{n_{>}-n_{<}}\left(  \frac{l_{B}^{2}k_{\bot}^{2}%
}{2}\right)  \right]  ^{2}\zeta(n,n^{\prime},\sigma,k_{z}). \label{S(k)}%
\end{align}
}}%
\begin{align}
\zeta(n,n^{\prime},\sigma,k_{z})  &  =2\kappa_{n_{>},\sigma}\Theta\left(
|k_{z}|\leq\kappa_{n_{<},\sigma}-\kappa_{n_{>},\sigma}\right) \nonumber\\
&  +(\kappa_{n_{>},\sigma}+\kappa_{n_{<},\sigma}-|k_{z}|)\Theta\left(
\kappa_{n_{<},\sigma}-\kappa_{n_{>},\sigma}\leq|k_{z}|\leq\kappa_{n_{>}%
,\sigma}+\kappa_{n_{<},\sigma}\right)
\end{align}
where $l_{B}=\sqrt{\frac{\hbar c}{eB}}$ is the magnetic length, $k_{\perp}$
denotes the component of the wave vector perpendicular to the applied magnetic
field, $n_{>}\equiv\max(n,n^{\prime})$ and $n_{<}\equiv\min(n,n^{\prime})$.
I{n order to avoid numerical {{{{underflow or overflow in the recurrence
relations, it is important to }}}}consider renormalized{ associated} Laguerre
polynomials {%
\[
\mathcal{L}_{n-p}^{p}\left(  x\right)  =\sqrt{\frac{p!(n-p)!}{n!}}L_{n-p}%
^{p}\left(  x\right)  .
\]
}}}

It is clear from (\ref{S(k)}) that the dependence of $S(\mathbf{k})$ on the
wave vector component perpendicular to the magnetic field and on the wave
vector component along the magnetic field can be treated separately, and that
the introduction of a magnetic field results in an anisotropy of the structure factor.

In the present calculations, the electron density is described through the
parameter $r_{s}=r_{0}/a_{B}^{\ast}$, where $r_{0}=\left(  4\pi n_{0}%
/3\right)  ^{-1/3}$ is the Wigner-Seitz radius, and $a_{B}^{\ast}=\hbar
^{2}\varepsilon_{\infty}/\left(  e^{2}m\right)  $ is the effective Bohr
radius. 

\begin{figure}
[ptbh]
\begin{center}
\includegraphics[
height=2.284in,
width=5.047in
]%
{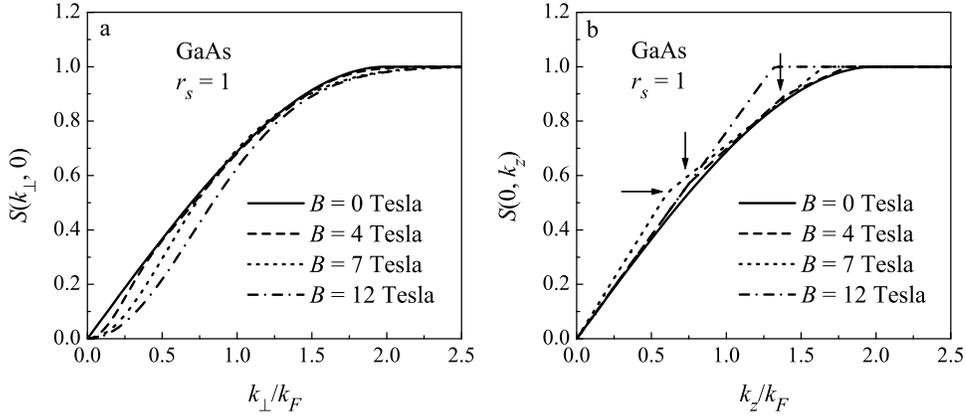}%
\caption{Hartree-Fock static structure factor ${\normalsize S}\left(
k_{\perp},k_{z}\right)  $ of an electron gas as a function of the wavevector
component $k_{\perp}$ perpendicular to the applied magnetic field (a) and as a
function of the wavevector component $k_{z}$ along the applied magnetic field
(b). In both cases the other wavevector component is equal to zero. In the
panel \textquotedblleft b\textquotedblright, the downward pointing arrows
indicate the two kinks corresponding to the magnetic field of 4 Tesla
($n_{\max}=2$), and the horizontal pointing arrow indicates the kink
corresponding to 7 Tesla ($n_{\max}=1$).}%
\label{fig:Sq}%
\end{center}
\end{figure}

In Fig. \ref{fig:Sq}, the static structure factor $S\left(  k_{\perp}%
,k_{z}\right)  $ is plotted for the value $r_{s}=1$, corresponding to the
electron density $n_{0}\approx3.75\times10^{17}$cm$^{-3}$ in GaAs. The
dependence of the structure factor on {$k_{\perp}$} is shown in Fig.
\ref{fig:Sq}(a), which reveals that the structure factor $\left.  S\left(
k_{\perp},k_{z}\right)  \right\vert _{k_{z}=0}$ is a smooth function with
limiting values $S\left(  0,0\right)  =0$ and $S\left(  \infty,0\right)  =1.$
The curvature however increases with increasing magnetic field, with a smaller
slope for small $k$, and an increasing value of the wave vector before the
high wave vector limit is reached.%

%TCIMACRO{\FRAME{ftbphFU}{5.047in}{2.284in}{0pt}{\Qcb{Hartree-Fock static
%structure factor $\QTR{normalsize}{S}\left(  k_{\perp},k_{z}\right)  $ of an
%electron gas as a function of the wavevector component $k_{\perp}$
%perpendicular to the applied magnetic field (a) and as a function of the
%wavevector component $k_{z}$ along the applied magnetic field (b). In both
%cases the other wavevector component is equal to zero. In the panel
%\textquotedblleft b\textquotedblright, the downward pointing arrows indicate
%the two kinks corresponding to the magnetic field of 4 Tesla ($n_{\max}=2$),
%and the horizontal pointing arrow indicates the kink corresponding to 7 Tesla
%($n_{\max}=1$).}}{\Qlb{fig:Sq}}{polgasgtekp-fig1r.eps}%
%width 5.047in;  height 2.284in;  depth 0pt;  original-width 3.5682in;
%original-height 1.6042in;  cropleft "0";  croptop "1";  cropright "1";
%cropbottom "0";  filename 'polgasGTEKP-Fig1r.eps';file-properties "XNPEU";}}}%
%BeginExpansion
%EndExpansion

Fig. \ref{fig:Sq}(b) reveals that the structure factor as a function of the
wave vector component along the magnetic field is far less smooth.
{\normalsize Relatively pronounced kinks appear in }$S\left(  0,k_{z}\right)
$ {\normalsize where the slope suddenly changes, and the number of kinks for
}$S\left(  \mathbf{k}\right)  <1$ {\normalsize exactly equals the number of
fully occupied Landau levels. The static structure factor }$S(k)$
{\normalsize thus allows to count the number of occupied Landau levels of a
polaron gas with given density and magnetic field strength. Furthermore,
}$S\left(  k_{\perp},k_{z}\right)  $ {\normalsize more rapidly tends to unity
in the }$z${\normalsize -direction, parallel to }$B${\normalsize , than in the
}$xy${\normalsize -plane. Furthermore, this anisotropy enhances with
increasing }$B${\normalsize . As a result, }$S\left(  k_{\perp},0\right)  $
{\normalsize is, as a general trend, a decreasing function of }$B,$
{\normalsize whereas }$S\left(  0,k_{z}\right)  $ {\normalsize increases with
increasing }$B.$ {\normalsize The physical reason for this anisotropy is the
following. A magnetic field applied along the }$z${\normalsize -axis hinders
the translation of the electrons in the }$xy${\normalsize -plane analogously
to a parabolic confinement potential with the frequency }$\omega_{c}$
{\normalsize and with confinement length }$l_{B}=\sqrt{\frac{\hbar c}{eB}}%
${\normalsize . The magnetic length decreases with increasing magnetic field
strength, and hence the region of }$k_{\perp}${\normalsize , in which
}$S\left(  k_{\perp},k_{z}\right)  $ {\normalsize substantially varies,
increases because it is of the order of }$l_{B}^{-1}${\normalsize , as seen
from the factor }$\exp\left(  -\frac{l_{B}^{2}k_{\bot}^{2}}{2}\right)  $
{\normalsize in Eq. (\ref{S(k)}). However, the anisotropy is non-monotonous:
it is accompanied by oscillations due to the de Haas -- van Alphen effect
\cite{KittelQTS}.}

\section{Ground state energy: results and discussion}

\label{GSE}

Converting the sums in (\ref{grondperdeel}) to integrals over cylinder
coordinates, introducing dimensionless variables, reducing the singularity in
the origin and using two-dimensional open Romberg integration, we calculated
the variational ground-state energy per particle numerically.%

%TCIMACRO{\FRAME{ftbphFU}{5.047in}{2.2096in}{0pt}{\Qcb{The ground-state energy
%per particle (a) and the Fermi energy (b) as a function of the magnetic field
%for a polaron gas in GaAs with $r_{s}=1$. The arrows and numbers indicate the
%changes of the number of filled Landau levels $n_{\max}$. The dashed curve
%shows the ground-state energy calculated without the polaron contribution.
%Notice the different scale in the abscises of both panels.}}{\Qlb{fig:Ers3}%
%}{polgasgtekp-fig2r.eps}{\special{ language "Scientific Word";
%type "GRAPHIC";  maintain-aspect-ratio TRUE;  display "USEDEF";
%valid_file "F";  width 5.047in;  height 2.2096in;  depth 0pt;
%original-width 3.5284in;  original-height 1.5342in;  cropleft "0";
%filename 'polgasGTEKP-Fig2r.eps';file-properties "XNPEU";}}}%
%BeginExpansion
\begin{figure}
[ptbh]
\begin{center}
\includegraphics[
height=2.2096in,
width=5.047in
]%
{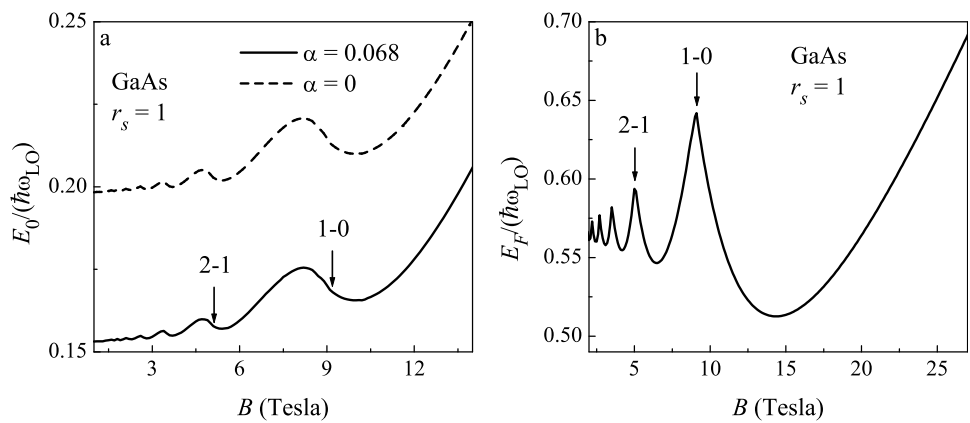}%
\caption{The ground-state energy per particle (a) and the Fermi energy (b) as
a function of the magnetic field for a polaron gas in GaAs with $r_{s}=1$. The
arrows and numbers indicate the changes of the number of filled Landau levels
$n_{\max}$. The dashed curve shows the ground-state energy calculated without
the polaron contribution. Notice the different scale in the abscises of both
panels.}%
\label{fig:Ers3}%
\end{center}
\end{figure}
%EndExpansion

In Fig. \ref{fig:Ers3}(a) the ground-state energy per particle is plotted as a
function of the magnetic field. Both the many-polaron ground-state energy and
the Fermi energy are oscillating functions of the magnetic field. At
well-defined magnetic fields inflection points occur in $E_{0}\left(
B\right)  $. This phenomenon can be understood if we look at the Fermi energy:
the inflection points of the ground-state energy coincide with the local
maxima of the Fermi energy, and therefore those inflection points are a direct
consequence of the de Haas -- van Alphen effect. {\normalsize For comparison,
we also have plotted in Fig. \ref{fig:Ers3}(a) the ground-state energy
calculated for }$\alpha=0$ {\normalsize (the dashed curve), i.~e., without the
polaron contribution. As seen from the figure, the electron-phonon interaction
substantially shifts the ground-state energy to lower values with respect to
that calculated for }$\alpha=0${\normalsize . This shift is practically
independent on the magnetic field strength. }

\begin{figure}
[ptbh]
\begin{center}
\includegraphics[
height=2.2528in,
width=5.0462in
]%
{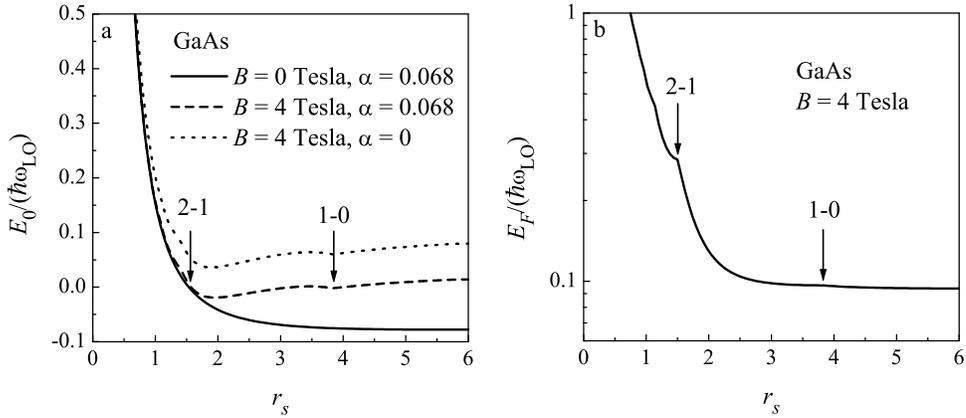}%
\caption{The ground-state energy per particle with and without applied
magnetic field (a) and the Fermi energy for $B=4$ T (b) for a polaron gas in
GaAs as a function of the parameter $r_{s}$. The arrows and numbers indicate
the changes of $n_{\max}$. The dotted curve shows the ground-state energy
calculated without the polaron contribution.}%
\label{fig:EB30}%
\end{center}
\end{figure}

The described dependence of the energy on the magnetic field is also seen in
Fig. \ref{fig:EB30}(a), which shows the energy as a function of the parameter
$r_{s}$. From this figure, it is clear that by applying a magnetic field, the
minimum of the ground-state energy occurs at a finite value of $r_{s}$.%

%TCIMACRO{\FRAME{ftbphFU}{5.0462in}{2.2528in}{0pt}{\Qcb{The ground-state energy
%per particle with and without applied magnetic field (a) and the Fermi energy
%for $B=4$ T (b) for a polaron gas in GaAs as a function of the parameter
%$r_{s}$. The arrows and numbers indicate the changes of $n_{\max}$. The dotted
%curve shows the ground-state energy calculated without the polaron
%contribution.}}{\Qlb{fig:EB30}}{polgasgtekp-fig3r.eps}%
%width 5.0462in;  height 2.2528in;  depth 0pt;  original-width 3.5561in;
%original-height 1.5766in;  cropleft "0";  croptop "1";  cropright "1";
%cropbottom "0";  filename 'polgasGTEKP-Fig3r.eps';file-properties "XNPEU";}}}%
%BeginExpansion
%EndExpansion
Similarly as in Fig. \ref{fig:Ers3}(a), kinks appear due to the change of the
number of fully occupied Landau levels, as is more clearly illustrated in Fig.
\ref{fig:EB30}(b) where the Fermi energy is shown as a function of $r_{s}$.
For all values of the magnetic field, the magnitude of the difference between
$E^{0}$ and $\left.  E^{0}\right\vert _{B=0}$ falls down with increasing
density, so that the influence of the magnetic field on the ground-state
energy of a polaron gas is larger for lower densities. This behavior of
$E^{0}$ as a function of the density is explained by the fact that, when the
density is high enough, a large number of Landau levels are filled. In this
case, the magnetic field should be considered as relatively weak with respect
to the characteristic frequency of the electron gas $E_{F}/\hbar$ ($\omega
_{c}\ll E_{F}/\hbar$).

{\normalsize When comparing in Fig. \ref{fig:EB30} the density dependence of
the ground-state energy for a polaron gas with that calculated without the
electron-phonon interaction, we see that the magnitude of the polaron
contribution decreases with decreasing }$r_{s}$ {\normalsize due to screening.
The same trend was indicated also for a polaron gas in the absence of a
magnetic field in Ref. \cite{LDB77}.}

\section{Conclusion}

We have extended the variational approach to the many-polaron problem
\cite{LDB77} to a system of interacting polarons in an external homogeneous
magnetic field. The rigorous upper bound for the ground-state energy of an
interacting polaron gas in a magnetic field has been obtained in terms of the
static structure factor. Due to the presence of a magnetic field, the static
structure factor as a function of the electron density reveals features
related to the filling of the Landau levels, when the Fermi energy crosses a
Landau level. The other result of an applied magnetic field is the
field-induced anisotropy of the structure factor $S\left(  \mathbf{k}\right)
$, which should be observable experimentally.

The application of a magnetic field has pronounced effects on the behavior of
the ground-state energy as a function of the electron density. The most
prominent feature, which is also observed in the ground-state energy as a
function of the magnetic field, is the occurrence of kinks in the ground-state
energy, which occur as a consequence of the de Haas -- van Alphen effect as
soon as a Landau level becomes fully occupied. Although the ground-state
energy is not directly observable, these phenomena should influence
experimental data, e.g., the optical absorption of a polaron gas in a magnetic
field, which are currently being studied.

\textbf{Acknowledgement.} This work has been supported by the IAP
(Interuniversity Attraction Poles), FWO-V projects G.0306.00, G.0274.01N,
G.0435.03, the WOG WO.025.99N (Belgium) and the European Commission GROWTH Programme.

\end{document}